\documentclass[twocolumn,floatfix,showpacs,preprintnumbers,amsmath,amssymb]{revtex4}
\usepackage{graphicx}
\usepackage{epstopdf}
\usepackage{dcolumn}
\usepackage{bm}
\usepackage[dvips]{color}

\begin{document}

\title{Roton Instabilities and Wigner Crystallization of Rotating Dipolar Fermions in the Fractional
       Quantum Hall Regime}

\author{Shih-Da Jheng$^{1}$}
\author{T. F. Jiang$^{1}$}
\author{Szu-Cheng Cheng$^{2,}$}
\email {sccheng@faculty.pccu.edu.tw}
\thanks{FAX: +886-2-28610577}

\affiliation{$^{1}$Institute of Physics, National Chiao Tung University, Hsinchu 30010, Taiwan, R. O. C. \\
             $^{2}$Department of Physics, Chinese Culture University, Taipei 11114, Taiwan, R. O. C.}


\date{\today}

\begin{abstract}
We point out the possibility of occurring instabilities in Laughlin liquids of rotating dipolar fermions with zero thickness.  Previously such a system was predicted to be the Laughlin liquid for filling factors $\nu \geq 1/7$.  However, from intra-Landau-level excitations of the liquid in the single-mode approximation, the roton minima become negative and Laughlin liquids are unstable for $\nu \leq 1/7$. We then conclude that there are correlated Wigner crystals for $\nu \leq 1/7$.
\end{abstract}


\maketitle

The cold quantum gas with the dipole-dipole interaction (DDI) was first realized in Cr atoms \cite{1}.  The quantum gases with the DDI are qualitatively different from non-dipolar ones \cite{2}. The novel anisotropic and long-range nature of the DDI offers a broad range of strong correlated many-body physics \cite{3,4}.  New quantum phases were predicted for the dipolar Bose-Einstein condensate (BEC) \cite{5, 6}.  The influence of the trapping geometry on the stability of the BEC and the effect of the DDI on the excitation spectrum were studied \cite{7}.  The vortex lattice of the rotating dipolar BEC exhibits novel bubble, stripe, and square structures \cite{8}.  For the dipolar Fermi gases, the \emph{s}-wave scattering is prohibited due to the Pauli Exclusion Principle and the Fermi surface is distorted by dipolar effects \cite{9}.   Bond pairs of fermions with resonant interaction are formed and the system of a Fermi gas behaves as a bosonic gas of molecules \cite{10, 11}.  The observed pairing of fermions provides the crossover between the weakly-paired, strongly overlapping Bardeen-Cooper-Schrieffer regime, and the tightly bound, weakly-interacting diatomic molecular BEC regime \cite{12}.  The strong correlations of fermions induced by the DDI can then be explored, such as the dipolar-induced superfluidity \cite{13, 14} and fractional-quantum-Hall-effect (FQHE) states in rotating dipolar Fermi gases \cite{15, 16}.

Rotating gases feel the Coriolis force in the rotating frame.  The Coriolis force on rotating gases is identical to the Lorentz force of a charged particle in a magnetic field.  Quantum-mechanically, energy levels of a charged particle in a uniform magnetic field show discrete Landau levels.  In the lowest Landau level (LLL), the kinetic energy of rotating dipolar gases is frozen and the DDI create strong correlations on particles.  Therefore, in the lowest Landau level, the potential energy from DDI dominates the kinetic energy and rotating dipolar fermions will crystallize into a Wigner crystal (WC) or become the FQHE liquid.  Baranov \emph{et al}. \cite{16} have shown that the FQHE states have lower energies than the WCs as filling factors $\nu \geq 1/7$ and in the zero-extension limit along the rotating axial direction, where $\nu=2\pi \rho \ell^2$.  Here $\rho$ and $ \ell$ are the average density and the magnetic length, respectively.  Although they also investigated the stability of the WC and found that the WC states were stable in the regime $\nu < 1/7$, but there was no test on the stability of the FQHE liquid.  It is still a question whether the FQHE liquid for rotating dipolar fermions is stable.

The phase transition point of the WC and FQHE states is usually determined by comparing their energies.  But this is not enough for a consistent theoretical description of the FQHE.  A consistent theory should tell us when the FQHE will not occur and becomes unstable \cite{17}.  One possible signature for instability can be revealed from the softening of the roton gap \cite{18}, which is an energy minimum at finite wave vector of the collective excitations of the FQHE state.  While the roton gap decreases with $\nu$, the collective excitations of FQHE states for a two-dimensional (2D) electron gas do not show any instability at the small $\nu$ regime where the actual ground state is the WC \cite{17,18,19}.  Such a shortcoming of the theory of the FQHE is repaired by applying the composite-fermion scheme to observe the collapse of the roton gap of the FQHE liquid at small $\nu$ \cite{17}.  For the present theoretical description of rotating dipolar fermions, it is not clear whether the excitation gap do cross zero at small $\nu$.  It would be interesting to look for the similar instability, which may provide a clue to the true ground states, in rotating dipolar fermions.

\begin{table*}[t]
\tabcolsep=18pt
\caption{Coefficients of the radial distributions of FQHE states.}
\begin{tabular}[c]{cccccccc}
  \hline
  $ $ & $C_1$ & $C_3$ & $C_5$ & $C_7$ & $C_9$ & $C_{11}$ & $C_{13}$  \\
  \hline
  $\nu$=1/3 & -1 &  17/32 &  1/16 & -3/32  &  0     &  0     &  0  \\
  $\nu$=1/5 & -1 & -1     &  7/16 &  11/8  & -13/16 &  0     &  0  \\
  $\nu$=1/7 & -1 & -1     & -1    & -25/32 &  79/16 & -85/32 &  0  \\
  $\nu$=1/9 & -1 & -1     & -1    & -1     & -29/8  &  47/4  & -49/8  \\
  \hline
\end{tabular}
\end{table*}

In this letter we report the first evaluation of intra-Landau-level excitation energies of FQHE states in rotating 2D polarized dipolar gases whose mass is \emph{M}.  Dipolar gases are fully polarized and their extension is zero in the rotating axial direction.  The approach is similar to the single-mode approximation (SMA) used in calculations of intra-Landau-level excitation energies of FQHE states in a 2D electron gas at high magnetic field \cite{18}.  We focus on low-energy excitations of the density oscillations in the extreme quantum limit where the ground state of a FQHE state is formed entirely within the LLL.  To avoid the inhomogeneous effect in the 2D direction, we studied the system rotating in the limit of critical rotation, where the magnitude of the rotating frequency $\Omega$ of the system is close to but still smaller than the trapping frequency.  Under the limit of critical rotation, the density of the trapped gas becomes uniform except at the boundary given by a trapping potential.  The excitation spectra of FQHE states exhibit a finite excitation-energy gap at \emph{k}=0 and a roton minimum at finite wave vector. The roton minimum decreases with $\nu$ and surprisingly becomes negative when $\nu \leq 1/7$, indicating an absence of FQHE state.  Note that the ground state of rotating dipolar fermions at $\nu=1/7$ is supposed to be the FQHE state shown by Baranov \emph{et al}. \cite{16}. We conclude that the Hartree-Fock WC state does not provide the correct estimate of the ground state energy at $ \nu= 1/7$. There exists a strong correlation effect on the Wigner crystallization of rotating dipolar gases at small $\nu$.

The formal development of mathematics within the subspace of the LLL has been presented elsewhere \cite{20}.  We shall apply previous formalisms and take the magnetic length $a=\sqrt{\hbar/2M\Omega}=1$.  The projected density operator with \emph{N} particles within the LLL is given by \cite{18, 20}
   \begin{equation}
   \bar{\rho}_\textbf{k}=\sum^N_{j=1}\text{exp}(-ik\partial/\partial z_j)\text{ exp}(-ik^{\ast}z_j/2),
   \end{equation}
where $z_j=x_j+iy_j$ and $k=k_x+ik_y$ are the complex representations of the \emph{j}th particle position and the wave vector of the density oscillations.  The projected density operators satisfy a commutation relation defined by \cite{18}
   \begin{equation}
   [\bar{\rho}_\textbf{k},\bar{\rho}_\textbf{q}]=\left(e^{k^{\ast}q/2}-e^{kq^{\ast}/2}\right)\bar{\rho}_{\textbf{k}+\textbf{q}}.
   \end{equation}
We are now equipped with tools to project the Hamiltonian into the LLL from the definition of the projected density operator.  The kinetic energy is constant and neglected due to the LLL approximation.  The projected Hamiltonian is written as
   \begin{equation}
   H=\frac{1}{2(2\pi)^2} \int d^2\textbf{q } V(\textbf{q}) \left(\bar{\rho}_{-\textbf{q}} \bar{\rho}_{\textbf{q}} -\rho e^{-q^2/2} \right),
   \end{equation}
where $V(\textbf{q})=-2\pi qD/a^3$ is the Fourier transform of the 2D dipolar-interaction potential with the coupling constant \emph{D}.

Following the previous SMA theory on the evaluation of excitation energies of the FQHE [18], we construct the collective-excitation state $|\textbf{k}\rangle$ with momentum $\textbf{k}$ as $|\textbf{k}\rangle
=N^{-\frac{1}{2}}\bar{\rho}_\textbf{k} |\text{0}\rangle $, where \emph{N} is the particle number and $|\text{0}\rangle$ is the
ground state with energy $E_0$. From $H|\textbf{k}\rangle =E_\emph{k}|\textbf{k}\rangle$, we can evaluate the
excitation energy $\Delta(k)=E_k-E_0$ by the formula $\Delta(k)=\bar{f}(k)/\bar{s}(k)$, where $\bar{f}(k)$
and $\bar{s}(k)$ are the projected oscillator strength and static structure factor, respectively.  Using Eq. (3),
$\bar{f}(k)= \langle 0| [\bar{\rho}_{-\textbf{k}},[H,\bar{\rho}_{\textbf{k}}]]|0\rangle/2N$ is readily evaluated with
the commutation relation given by Eq. (2) and we have
   \begin{eqnarray}
   \bar{f}(k)&=&\frac{1}{2}\sum_{\textbf{q}} V(\textbf{q})\left(e^{q^{\ast}k/2}-e^{qk^{\ast}/2}\right) \times \nonumber \\
   &&\left[\bar{s}(q) e^{-k^2/2} \left(e^{-qk^{\ast}/2}-e^{-q^{\ast}k/2}\right) \right. \nonumber \\
   &&+\left. \bar{s}(k+q)\left(e^{qk^{\ast}/2}-e^{q^{\ast}k/2}\right) \right].
   \end{eqnarray}
The projected static structure factor is related to the Fourier transform of the radial distribution \emph{g}(\textbf{r}) of the ground state, $\bar{s}(k)=\rho\int d^2\textbf{r } \text{exp}(-i\textbf{k}\cdot \textbf{r})g(\textbf{r})+\text{exp}(-k^2 /2)$, where
   \begin{equation}
   \emph{g}(\textbf{r})=1-e^{-r^2/2}+\sum^{\infty}_{m=0} \frac{C_{2m+1}}{(2m+1)!} \left(r^2/4\right)^{2m+1} e^{-r^2/4}.
   \end{equation}
The coefficients $C_{2m+1}$ were derived by the sum rules from the physical properties of Laughlin's wave functions \cite{19} and shown in Table 1 for different fractional filling factors. $C_{2m+1}$=0 for $m\geq7$.  Having obtained an analytic form of the radial function, ground-state energies and the projected static structure factors of FQHE states are readily evaluated. For the zero extension in the rotating axial direction, the ground-state energies $E_0(\nu)$ for $\nu=1/3$, 1/5, 1/7 and 1/9 are $0.3690(1/3)^{3/2}D/a^3$, $0.3361(1/5)^{3/2}D/a^3$, $0.3228(1/7)^{3/2}D/a^3$ and $0.3105(1/9)^{3/2}D/a^3$ respectively.  These energies are in excellent agreement with the energies from the Monte Carlo method \cite{16}, where the ground-state energies are $0.3665(1/3)^{3/2}D/a^3$, $0.3348(1/5)^{3/2}D/a^3$, $0.3216(1/7)^{3/2}D/a^3$ and $ 0.3145(1/9)^{3/2}D/a^3$ for $\nu=1/3$, 1/5, 1/7 and 1/9, respectively.

\begin{figure}
\centering
\includegraphics[width=3.2in]{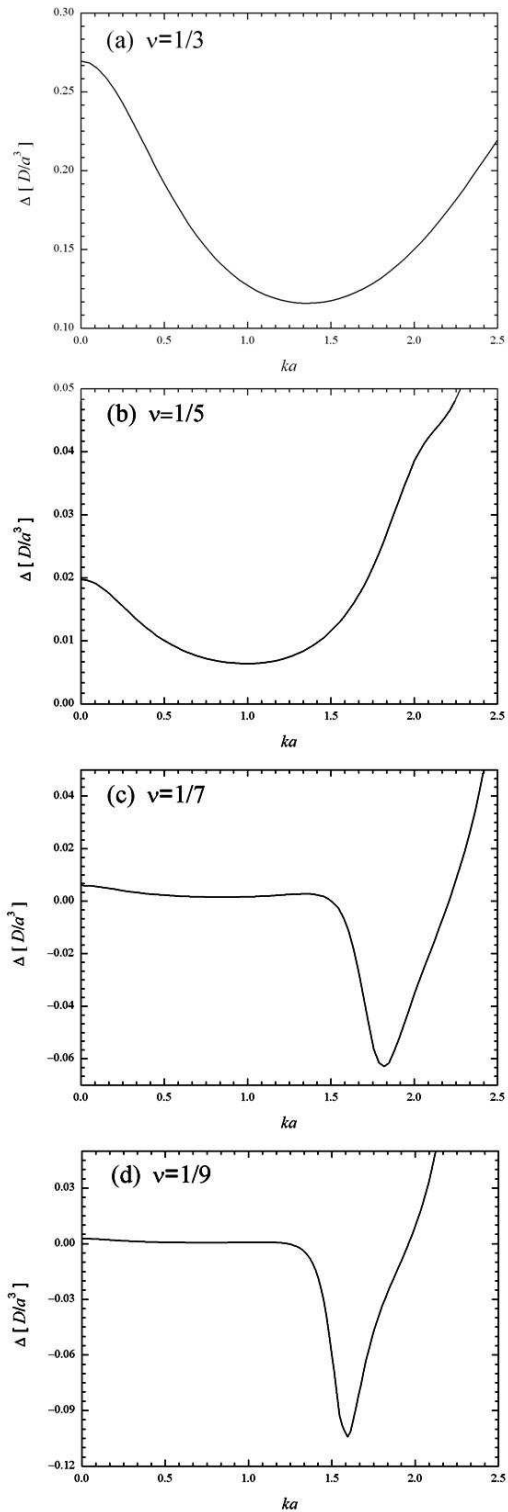}
\caption{Collective-excitation dispersions for FQHE states at $\nu=1/3$, 1/5, 1/7 and 1/9, respectively.  All curves are the results for the zero-extension limit in the rotating axial direction.}
\end{figure}

From the Fourier transform of Eq. (5), we compute $\bar{s}(k)$, then use this to evaluate $\bar{f}(k)$ and calculate the collective-excitation energy $\Delta(k)$ finally. That $\bar{s}(k) \sim k^4$ for small \emph{k} is a phenomenon of the lack of density fluctuations or the incompressibility of FQHE states at long wavelength.  This is the source of the finite energy gap observed in the FQHE of a 2D electron gas in a strong magnetic field.  Therefore, we compute $\Delta(k)$ at small \emph{k} using the exact leading term in $\bar{s}(k)$, which is $\bar{s}(k)=(1-\nu)k^4/8\nu$ in the long wavelength limit.  We studied the collective-mode dispersion for the zero-extension limit in the rotating axial direction.  The evaluated $\Delta(k)$ versus wave vector \emph{k} for FQHE states is shown in Fig. 1.  The essential features of excitation dispersions exhibit a finite excitation-energy gap at \emph{k}=0 and an energy minimum $\Delta$ (roton) at finite wave vector.  A finite excitation-energy gap at \emph{k}=0 implies that the density fluctuations cost energy and a rotating-dipolar-Fermi gas is an incompressible fluid.  The excitation-energy gap at \emph{k}=0 is decreasing with $\nu$.  The gap at \emph{k}=0 is rather small but remains positive at $\nu=1/9$, indicating that the FQHE is easily destroyed by the quantum fluctuations at very low $\nu$.

The roton structure of excitations at finite wave vector is caused by a peak in $\bar{s}(k)$, which is usually interpreted as a starting signal of crystallization [18].  The WC state is shown to be the ground state at very low $\nu$.  This fact is revealed by the roton energies being decreasing with $\nu$ in Fig. 1.  The collective-excitation dispersion of the roton has a finite energy gap for $\nu=1/3$ and 1/5, respectively, indicating that it cost finite energy at higher $\nu$ to make the FQHE state unstable.  This gap is lying close to the primitive reciprocal-lattice wave vector, \emph{G}, of a WC, which is given by $G_{1/3}a=1.555$ at $\nu=1/3$ and $G_{1/5}a=1.205$ at $\nu=1/5$, respectively.  There is a significant reduction in the roton minimum from $\nu=1/3$ to $\nu=1/5$.  The deepening of the roton gap in going from $\nu=1/3$ to $\nu=1/5$ is a signature of incipient crystallization near $\nu=1/7$.

If we lower $\nu$ further, the roton minimum is continuously decreasing and becomes negative as $\nu \leq 1/7$ (see Fig. 1(c) and 1(d)). This collapse of the gap shows that the FQHE state with excitations has lower energy than the FQHE state itself and no FQHE in the regime $\nu \leq 1/7$.  We know that no FQHE but the Wigner crystallization exists in the regime $\nu \leq 1/9$ [16], which is contradictory with our study.  We conclude from the roton instability that the phase boundary between the FQHE and WC states is at $\nu=1/7$.  Therefore, the most interesting thing of our study is that the FQHE states below $\nu=1/7$ is unstable to a spontaneous creation of excitations.  This unexpected instability suggests that there exists a strong correlation effect on the WC state which is ignored in the previous Hartree-Fock theory of the WC [16].  Note that the \emph{G}s of WC states at $\nu=1/7$ and 1/9 are given by $G_{1/7}a=1.018$ and $G_{1/9}a=0.898$, respectively.  Instead of near the \emph{G} of a WC, we find that the roton minima for $\nu=1/7$ and 1/9 are far from this wave vector and located at \emph{ka} = 1.833 and 1.583, respectively.  It seems that the collapse of the roton gap only indicates the instability of the FQHE state and does not show any signature of the Wigner crystallization at very low $\nu$.  In fact, we believe that one must introduce Laughlin-Jastrow correlations into the WC state to interpret the Wigner crystallization in terms of the softening of the roton minimum [21].  As shown in the reference 21, density fluctuations caused by displacements of particles from the lattice sites are suppressed by the Laughlin-Jastrow correlation effectively.  This delocalization of particles from the lattice sites makes the average lattice constant and \emph{G} of the WC becoming shorter and larger, respectively.  The fact that the wave vector, at which the roton minimum becomes negative, is far from the \emph{G} of the WC indicates the importance of the Laughlin-Jastrow correlation on the Wigner crystallization of dipolar fermions.

Having understood a physical picture of excitations, we now examine the validity of the SMA.  The SMA theory, being based on a variational ansatz, provides an energy upper bound to the lowest excited state.  For large \emph{k} the quasiparticle and the quasihole, which are a pair of quasiparticles created by the density-wave excitation, are far apart.  The upper bound of the excitation energy of the pair is given by the first excitation moment $\Gamma=2\left[\nu E_0(1)-E_0(\nu)\right]/(1-\nu)$ [18], where $E_0(1)$=0.6267$D/a^3$ is the ground-state energy at $\nu=1$.  Suppose the binding energy of a far apart quasiparticle-quasihole pair is very small due to the $1/r^3$ of the dipole-dipole interaction.  The upper bound of the excitation energies of a quasiparticle plus a quasihole is also provided by the moment $\Gamma$.  The moments  for $\nu=1/3$ and 1/5 are $\Gamma_{1/3}=0.4136D/a^3$ and $\Gamma_{1/5}=0.2382D/a^3$, respectively.  These values are above the roton gap, but lie considerably below the result of the quasiparticle excitation spectrum of Baranov \emph{et al}. [15]: the excitation energy of a quasihole is $\triangle\varepsilon_{qh}=0.9271D/a^3$ and the same order of magnitude is expected in the excitation energy of quasiparticles.  Although we do not have an explanation of this large discrepancy, we can certainly say that there are no low-lying excitations of quasiparticles and quasiholes below the roton mode.  The lack of low-lying single-particle excitations means that the dynamics of the system is completely described in terms of collective modes.  The validity of the SMA is proved.

We studied neutral excitations of rotating dipolar Fermi gases theoretically.  The possible experimental ways to verify theoretical results will be the next important issue.  One of the possible ways is using Bragg spectroscopy [22] to measure the excitation spectrum and the static structure factor of the FQHE.  Bragg spectroscopy of a Bose-Einstein condensate in a trap has been realized experimentally [23].  The measured excitation spectrum agreed well with the Bogoliubov spectrum for a condensate.  Hafezi \emph{et al}. [24] also proposed a detection technique based on Bragg spectroscopy to obtain the excitation spectrum and the static structure factor of the FQHE in an optical lattice.  Their study has shown that excitation gaps of the FQHE can be realized by Bragg spectroscopy.

In conclusion, we investigated intra-Landau-level excitations of a rotating quasi-2D dipolar system in the FQHE regime.  We show that the excitation gap at \emph{k}=0 and the roton minimum are decreasing with $\nu$.  We find that the roton minimum becomes negative and the FQHE states are unstable as $\nu\leq7$.  The wave vector at which the roton gap is collapsed is far from the primitive reciprocal-lattice wave vector of the WC.  This discrepancy indicates that one must introduce Laughlin-Jastrow correlations into the ground state of the WC to find the phase transition from the WC to the FQHE liquid.  We believe that this transition will occur between $\nu=1/5$ and $1/7$.

The author acknowledges the financial support from the National Science Council (NSC) of Republic of China under Contract No. NSC99-2112-M-034-002-MY3.  The author also acknowledges the support of the National Center for Theoretical Sciences of Taiwan during visiting the center.

\end{document}